\documentclass[journal=jacsat,manuscript=article,doi]{achemso}
\setkeys{acs}{articletitle=true,doi=true}
\usepackage[version=4]{mhchem} 
\usepackage{textcomp}
\usepackage{graphicx}
\usepackage{hyperref}
\usepackage{color}
\usepackage{soul}
\newcommand{\doi}[1]{\href{http://dx.doi.org/#1}{\nolinkurl{#1}}}

\author{Marco Cazzaniga}
\email{marco.cazzaniga@unimi.it}
\affiliation[ISTM]{Consiglio Nazionale delle Ricerche - Istituto di Scienze e Tecnologie Molecolari, via Golgi 19, 20133 Milano, Italy}
\altaffiliation{Present address: Department of Chemistry, University of Milan, via Golgi 19, 20133 Milano, Italy}

\author{Fausto Cargnoni}
\email{fausto.cargnoni@istm.cnr.it}
\affiliation[ISTM]{Consiglio Nazionale delle Ricerche - Istituto di Scienze e Tecnologie Molecolari, via Golgi 19, 20133 Milano, Italy}
\alsoaffiliation[INSTM]{Consorzio INSTM, UdR di Milano, Italy}

\author{Marta Penconi}
\affiliation[ISTM]{Consiglio Nazionale delle Ricerche - Istituto di Scienze e Tecnologie Molecolari, via Golgi 19, 20133 Milano, Italy}

\author{Alberto Bossi}
\affiliation[ISTM]{Consiglio Nazionale delle Ricerche - Istituto di Scienze e Tecnologie Molecolari, via Golgi 19, 20133 Milano, Italy}
\alsoaffiliation[INSTM]{Consorzio INSTM, UdR di Milano, Italy}

\author{Davide Ceresoli}
\email{davide.ceresoli@cnr.it}
\affiliation[ISTM]{Consiglio Nazionale delle Ricerche - Istituto di Scienze e Tecnologie Molecolari, via Golgi 19, 20133 Milano, Italy}
\alsoaffiliation[INSTM]{Consorzio INSTM, UdR di Milano, Italy}

\title{Unraveling the degradation mechanism of FIrpic based blue OLEDs: I. A theoretical investigation}

\begin{document}

\begin{tocentry}
  \begin{center}
  \includegraphics[width=8.47cm,height=4.76cm]{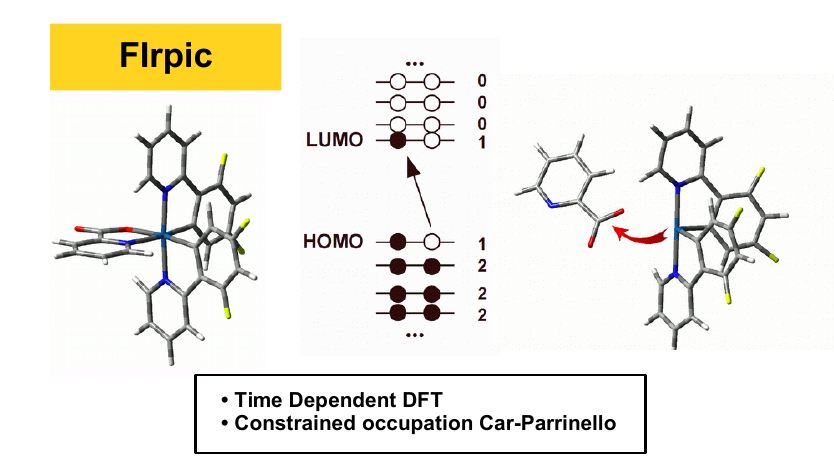}
  \end{center}
\end{tocentry}

\begin{abstract}
We report a detailed ab-initio study of the  of the microscopic
degradation mechanism of FIrpic, a popular blue emitter in OLED
devices. We simulate the \emph{operando} conditions of
FIrpic by adding an electron-hole pair (exciton) to the system.
We perform both static calculations with the TDDFT framework and
we also simulate the evolution of the system at finite temperature
via Car-Parrinello molecular dynamics. We found triplet excitons
are very effective in reducing the Ir-N bond breaking barrier of
the picolinate moiety. After the first bond breaking, the two
oxygen of picolinate swap their position and FIrpic can either
remain stable in an ``open'' configuration, or loose a picolinate
fragment, which at a later stage can evolve a CO$_2$ molecule.
Our method can be applied to other light emitting Ir-complexes
in order to quickly estimate their stability in OLED devices.
In Paper~II we complement our theoretical study with a parallel
experimental investigation of the key degradation steps of
FIrpic in an aged device.
\end{abstract}

\section{Introduction}
Organic light-emitting diodes (OLEDs) offer the potential of using both
the singlet and triplet excitons in realizing 100\% internal quantum
efficiencies of electro-luminescence (see Ref.~\cite{Minaev2014} for a
recent review). During the operation lifetime of a OLED device, the
injected charge carriers (electrons and holes) may become trapped at
morphological and chemical defects, and recombine non-radiatively.
These phenomena, not only limit the quantum efficiency of the device,
but are also responsible for the degradation of the device (leading
to aging and failure), through the formation of highly reactive radical
species~\cite{Scholz2015}. The danger of chemical degradation is
particularly enhanced in blue-emitting phosphorescent devices, because
of the large energy of the photon and of the long lifetime of the excited
triplet state.~\cite{rolloff} Several mechanism are responsible of the
degradation of the emission properties of PhOLEDS, either by quenching
emission, or by chemical degradation. These include triplet-triplet
annhilation (TTA), triplet-polaron annhilation (exciton-charge
interaction).~\cite{Sajoto2009,Forrest2015} These mechanism might not be
thermodynamically favorable, but when they take place, they can be
as detrimental as non-radiative exciton recombination.

In a real device, under operating conditions, chemical degradation occurs
both in the charge-transporting layers (\emph{host}) and in the emissive
layers, where \emph{guest} emitting dyes are dispersed. For instance
CBP (4,4$^\prime$-Bis(N-carbazolyl)-1,1$^\prime$-biphenyl), an efficient
hole-transporting material, can trap an extra charge (\emph{polaron}) that
can break one of the \ce{C-N} bonds. The resulting radical fragment is
highly reactive and will attach to a neighboring CBP molecule, modifying
dramatically its photophysical and transport properties.~\cite{Schmidbauer2013}
The degradation of FIrpic (Bis[2-(4,6-difluorophenyl)pyridinato-C$^2$,N](picolinato)Ir(III))
has been studied experimentally by Moraes and
coworkers~\cite{Moraes2011}, who performed mass spectroscopy experiments
on pristine and aged devices. From the observed fragments, the authors
were able to propose two degradation pathways: (i) detachement of the
picolinate; (ii) loss of a CO$_2$ molecule and possible sequential loss
of the pyridine. They also observed the formation of several Ir complex
incorporating ligands from the host molecules. Moreover the presence of
all FIrpic isomers was observed in aged devices~\cite{Citti2016}, despite
the fact that the synthetic route of FIrpic yields only one isomer.~\cite{Baranoff2015}

Density Functional Theory (DFT) and its time-dependent DFT (TDDFT)
calculations provide a wealth of useful insights into the photodeactivation,
inter-system crossing and non-radiative decay
processes.~\cite{Himmetoglu2012,Wang2014,Li2013,Escudero2015,Li2015,Xu2015,Escudero2015,Kesarkar2016,Jacquemin2017,Penconi2017,Penconi2018,Zhang2018,Jeong2019} 
Calculations on Ir(ppy)$_3$ and Ir(ppz)$_3$ showed that these complexes display
a stable ``open'' state, in which one of the the ligand break one bond with the
central Ir, followed by a rotation of the ligand along the remaining Ir-bond.
This open state is particularly favored in the triplet state.~\cite{Treboux2007}
The isomerization barriers of a small model Ir complex where calculated in
Ref.~\cite{Koseki2013}. However, the isomerization of larger, emissive Ir-complexes,
will require to overcome quite large energy barriers, because of the steric
hindrance of the ligands~\cite{ArroligaRocha2018}. Therefore, a complete
understanding of the degradation pathways, based on atomistic modeling, is still missing to date.

\begin{figure}
  \begin{center}
  \includegraphics[width=8cm]{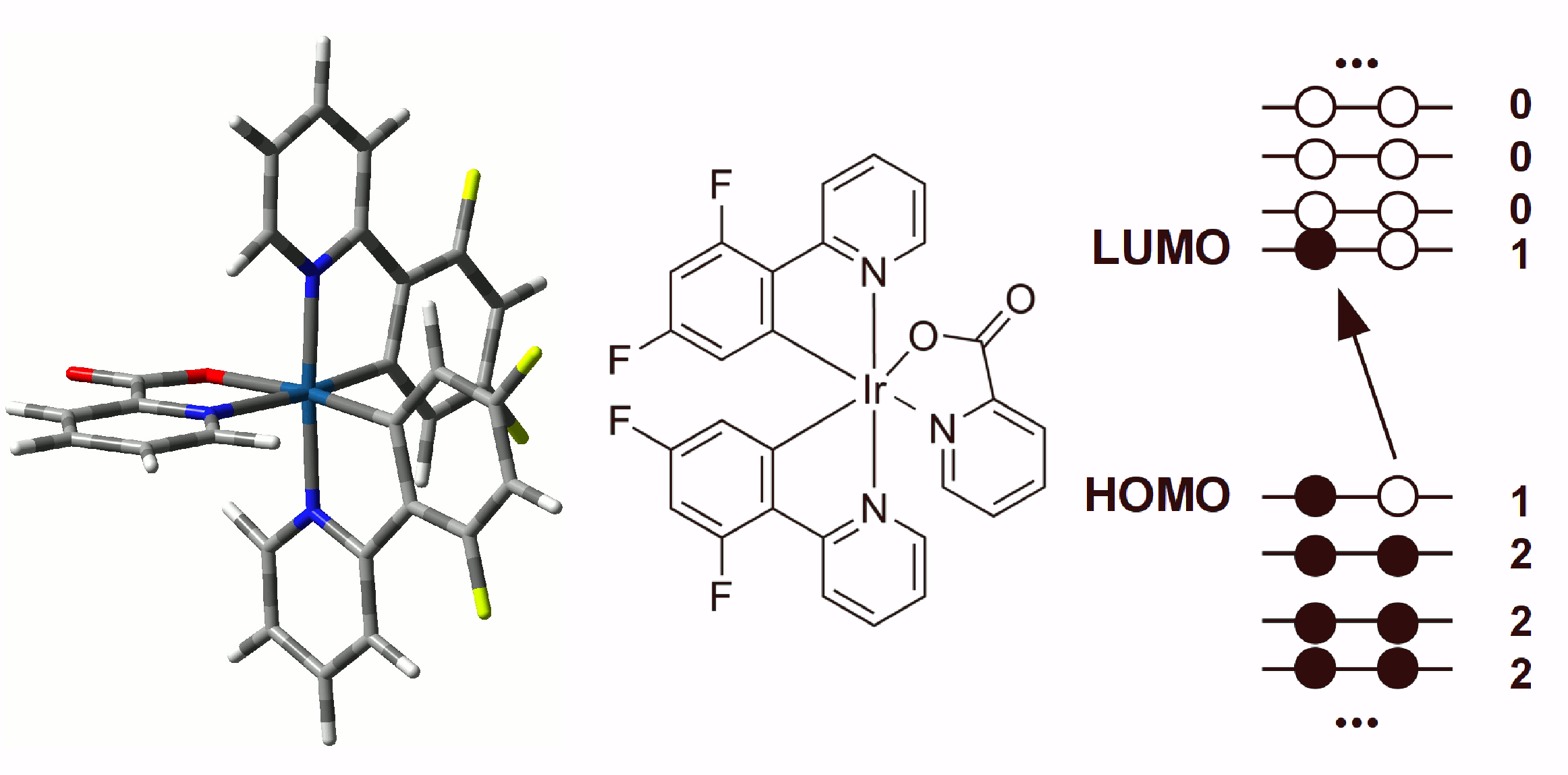}
  \caption{(Left panel) The most stable isomer of the closed FIpric molecule in
  its electronic ground state. Hydrogen atoms are painted in white, carbon in grey,
  nitrogen in blue, oxygen in red, fluorine in yellow, iridium in light blue.
  (Center panel) FIrpic chemical structure. (Right panel) Schematic representation
  of the orbital occupations of FIrpic during Car-Parrinello molecular dynamics
  runs in the excited state.}
  \label{fgr:firpic_closed}
  \end{center}
\end{figure}

In this manuscript we concentrate on the guest emitter FIrpic, whose most stable
isomer is shown in Fig.~\ref{fgr:firpic_closed}. To simulate a OLED \emph{operando}
conditions, it is necessary to perform DFT calculations in the charged and excited state.
In principle, excited states (singlet or triplet) can be modeled accurately only
by many-body perturbation theory such as the GW-BSE technique~\cite{Hedin65,Gunnarson98},
but this method is too computationally expensive even on modern supercomputers.
Moreover, non-adiabatic evolution of charges and excitons is usually simulated
by TDDFT Eherenfest dynamics~\cite{Li05} or Tully Surface Hopping~\cite{Tully71,Barbatti07}.
The time step needed for TDDFT propagation is of the order of attoseconds (10$^{-18}$~s),
and the simulation time is restricted to few picoseconds (10$^{-12}~$s), which
is sufficient to simulate fast radiative and non-radiative processes.

To simulate the degradation mechanisms, a much longer timescale ($\sim$10 picoseconds) is
needed, corresponding to the time occurring for molecular geometry rearrangements
and bond-breaking. Therefore, we perform Car-Parrinello molecular dynamics~\cite{Car1985}
with constrained orbital occupations (c-CPMD). This computational framework was used
successfully to simulate electron-polarons~\cite{Serra00b,Serra02}, exciton trapping
and chemical degradation in crystalline polyethylene.~\cite{Ceresoli04,Ceresoli05}
In those works on polyethylene, it was shown that c-CPMD method is accurate enough
to provide at least a qualitative picture of the degradation pathways. However, a
precise estimate of the probability and the timescale of the degradation process,
cannot be obtained with the c-CPMD method. We also performed a series of static
computations to simulate the initial step (i.e. bond breaking) of FIrpic. Interestingly,
in performing these calculations we discovered a possible mechanism of isomerization
of FIrpic requiring relatively low energy barriers.

\section{Computational details}
\subsection{Static computations}
We performend static computations (i.e. at zero temperature) of FIrpic in
three different electronic states: the singlet ground state (GS); the
first excited triplet (T$_1$), described both with an unrestricted DFT
scheme and the TDDFT method within the Tamm-Dancoff (TDA)
approximation~\cite{Hirata1999}; the first excited singlet (S$_1$),
adopting just the TDDFT-TDA level of theory. We employed both the PBE0 and
B3LYP hybrid functionals. The results obtained with the two functional
are slightly different but the overall picture of pathways for FIrpic
degradation is the same. In the next section we'll report only the
PBE0 results, while B3LYP results are reported the Supplementary Information
(SI).

We used the Gaussian03 package~\cite{Pople2003} and the 6-31g(d,p) basis
set for the light atoms (hydrogen, carbon, nitrogen, oxygen and fluorine),
and the aug-cc-pVDZ basis with 17 valence electrons for Ir. The remaining
60 core electrons were described with by a pseudopotential specifically
developed for this basis~\cite{Figgen2009}. Our basis set, which
includes diffuse as well as polarization functions, should be flexible
enough to ensure that the wavefunction might expand on an appropriate set
of Gaussian functions even in geometrical conformations far from the closed
isomer.

We considered four pathways possibly leading to FIrpic degradation:
breaking of Ir-O and Ir-N bonds of the picolinate fragment (pic), as well
as Ir-N and Ir-C bonds belonging to one of the (4,6-difluorophenyl)pyridinate
moieties (F$_2$ppy). In a single case we also detailed a complete
isomerization pathway through breaking of the bond between iridium
and the nitrogen atom belonging to the F$_2$ppy fragment. We tried to simulate
the isomerization of FIrpic simply by rotating one of the fragments, but
we always observed further bond breaking during the calculations.

All configurations presented here have been fully relaxed. The effect of
zero-point vibrational motion on the estimate of energy barriers and stability
of open conformations has been evaluated in selected geometries, and proved
almost negligible (i.e. of the order of 0.01~eV). This correction is
therefore not included in the data reported in the next section. Transition
states were determined using analytical derivatives in the case of DFT
single determinant computations, while TDDFT data came from numerical scans 
over the reaction coordinates, where all internal coordinates but one
were optimized.

\subsection{Dynamical simulations}
We carried out Car-Parrinello~\cite{Car1985} molecular dynamics simulations
of FIrpic imposing constrained occupations to simulate the presence of an
electronic excitation.
These simulations were performed with the
CPV code of Quantum~Espresso~\cite{Giannozzi2009,Giannozzi2017}.

We used ultrasoft pseudopotentials with a plane-wave energy cut-off of 45.6~Ry,
and the PBE functional. Isolated molecules are contained in a supercell with at
least 10~\AA\ of vacuum around the starting geometrical configuration.
Equation of motion of ions and electrons were integrated with the Verlet
algorithm, assigning to the electron a fictitious mass of 400 atomic units.
To control the dynamics we used a Nos\'e thermostat with a temperature of 1000~K for
ions, acting with a frequency of 90~THz. This same thermostat, acting with a frequency
of 110~THz, has been adopted for electrons in the simulations of polarons and
excitons, keeping as reservoir the asymptotic value of the electronic kinetic
energy obtained in the neutral-ground state computation. For each system we
performed 90,000 steps of molecular dynamics, which correspond to an overall time
evolution of $\sim$11~ps. A temperature as high as 1000~K might seem unrealistic,
and its sole purpose is to speed up the dynamical evolution. In any case, we
verified that such high temperature does not induce molecular fragmentation
in FIrpic, within the timescale of the simulation.

\section{Results and discussion}
\subsection{Static computations: Ring opening and isomerization}
\begin{figure}
  \begin{center}
  \includegraphics[width=8cm]{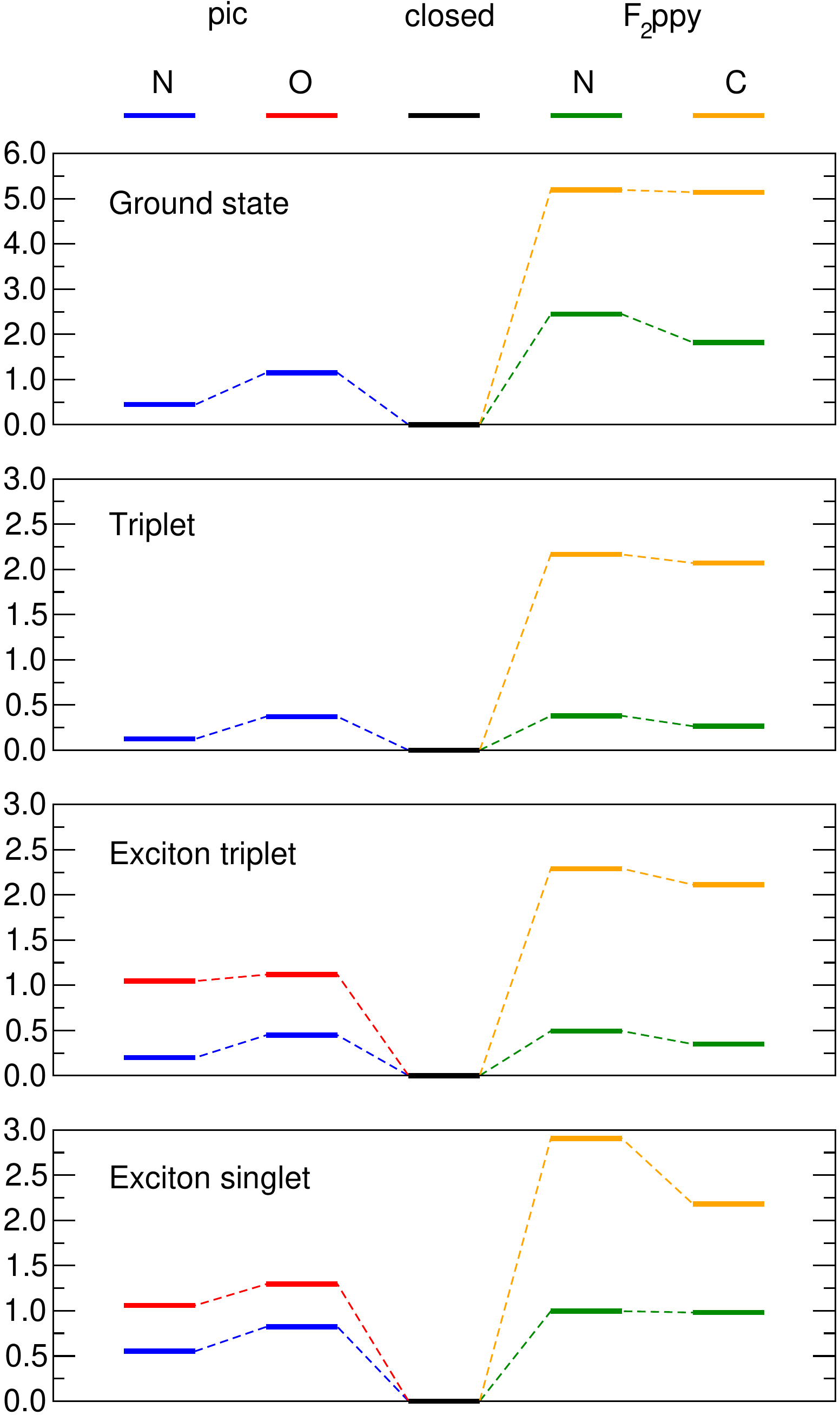}
  \caption{Relative energies (eV) of transition and metastable open states of
  FIrpic, determined for several electronic states (see caption in each panel).
  For convenience, the energy  of each closed conformation is set to
  zero. The opening of the picolinate fragment is reported on the left,
  while the opening of the (4,6-difluorophenyl)pyridinate (F$_2$ppy) is
  reported on the right. Every structure is fully relaxed. Data
  obtained with the PBE0 functional.}
  \label{fgr:plotstatespbe0}
  \end{center}
\end{figure}

\begin{figure}
  \begin{center}
  \includegraphics[width=14cm]{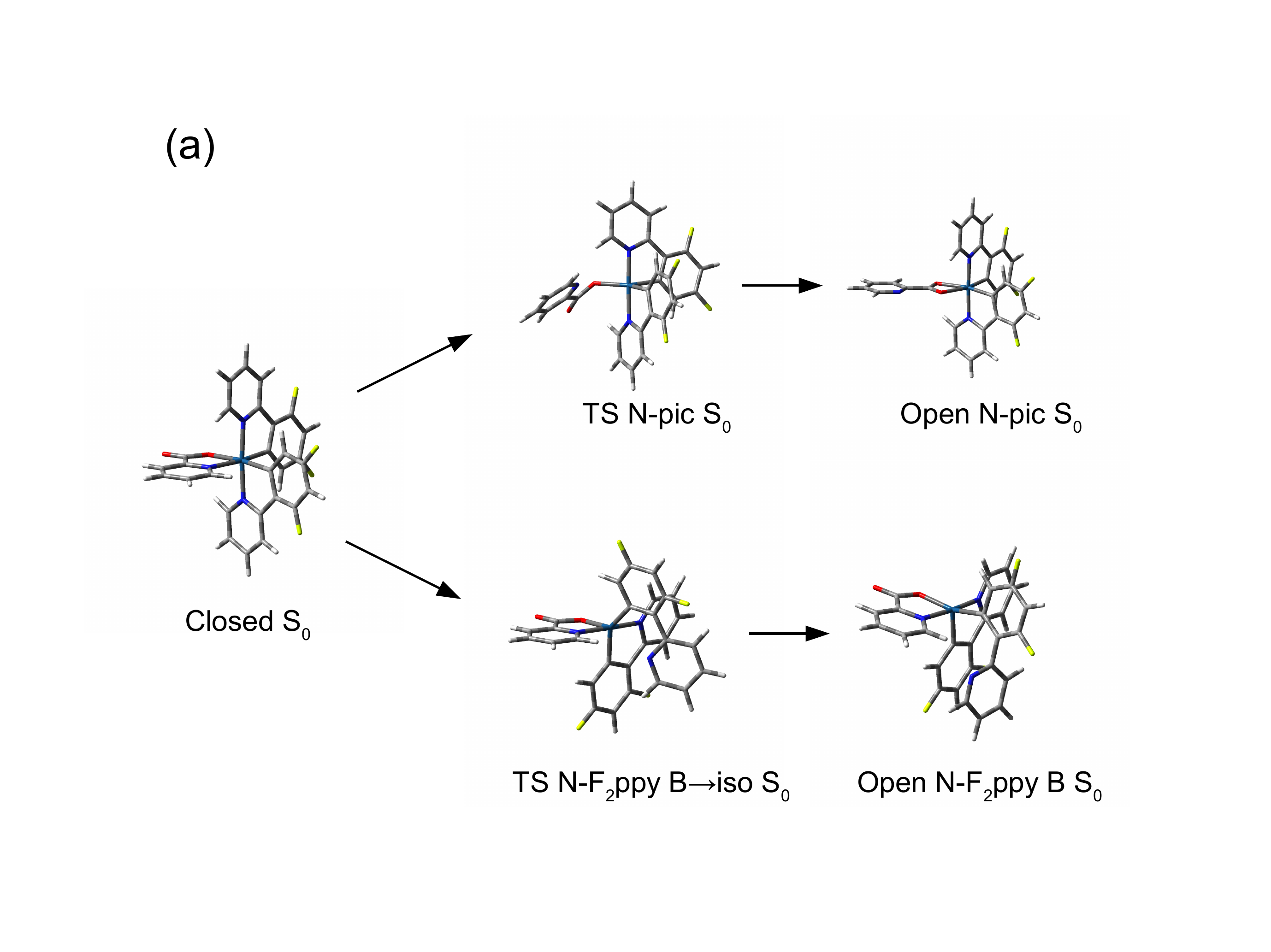}
  \includegraphics[width=14cm]{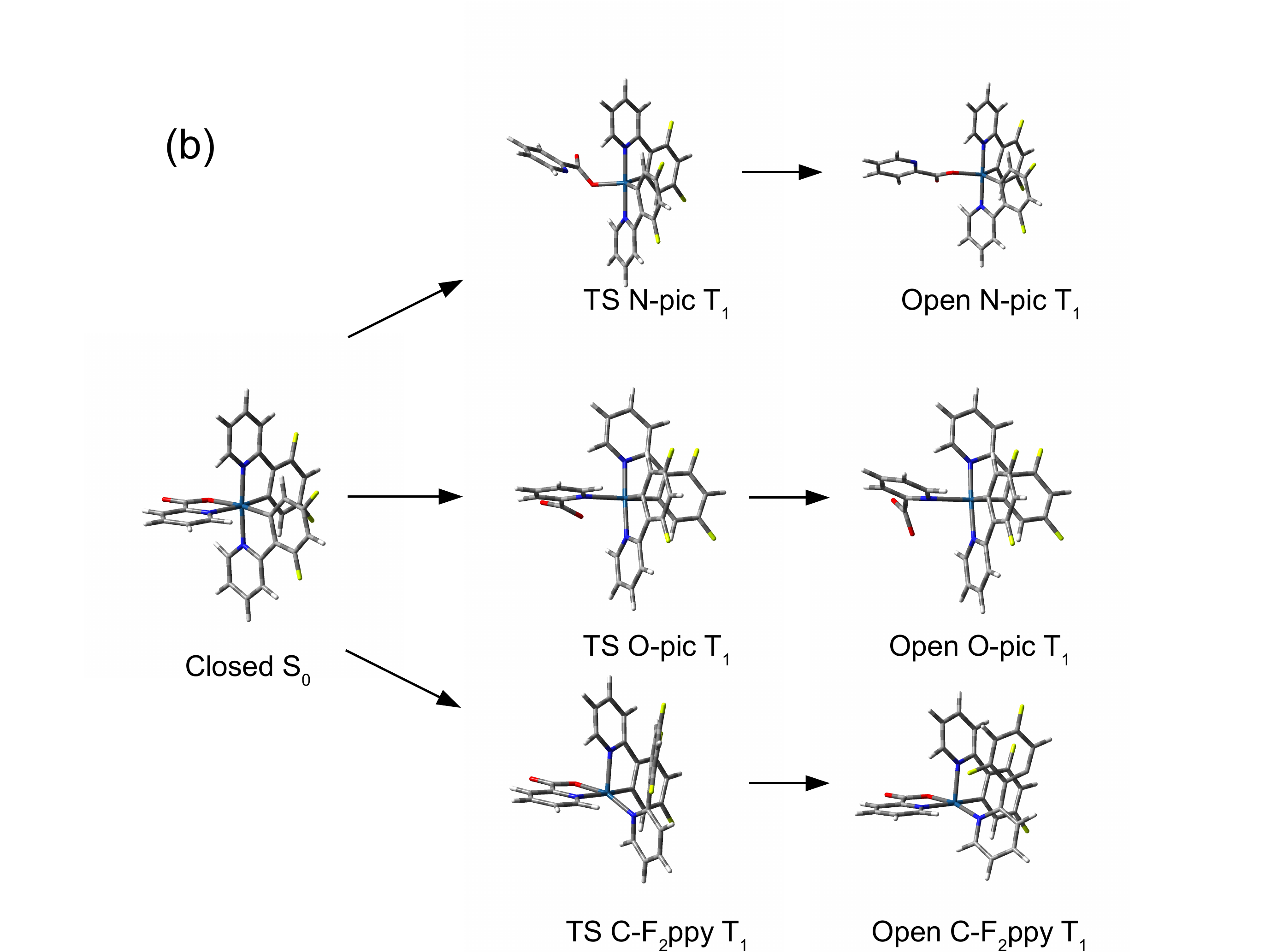}
  \caption{Selected conformations of FIrpic relevant to ligand opening.
  (a) Ir-N(pic) and Ir-N(F$_2$ppy) opening in the ground state (S$_0$).
  (b) Ir-N(pic), Ir-O(pic) and Ir-C(F$_2$ppy) opening in the excited triplet (T$_1$).
  All structures are fully relaxed. Data obtained with the PBE0 functional.}
  \label{fgr:Firpicopening}
  \end{center}
\end{figure}

\begin{figure}
  \begin{center}
  \includegraphics[width=14cm]{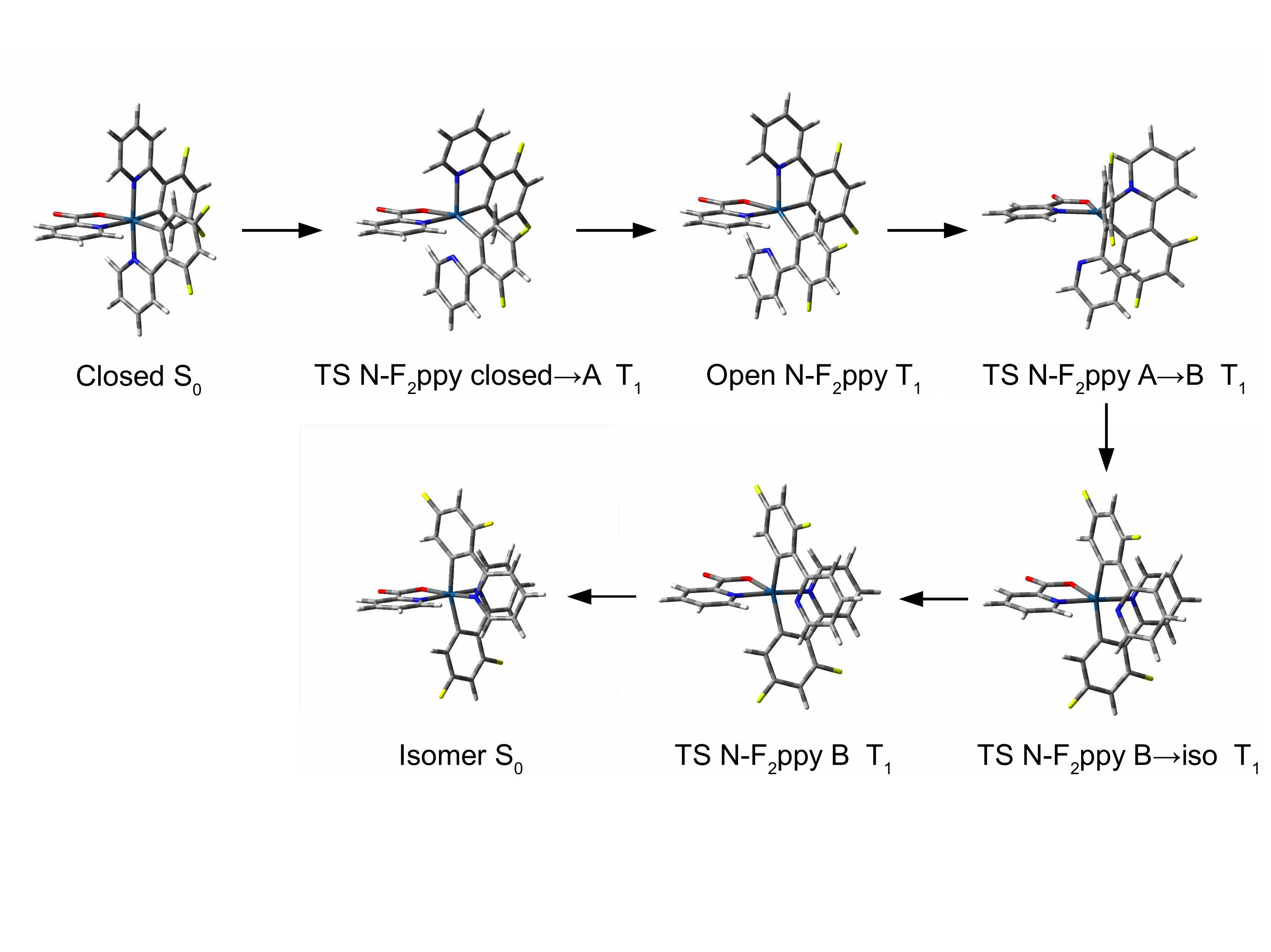}
  \caption{Selected conformations of FIrpic relevant to isomerization,
  in the T$_1$ state. All structures are fully relaxed. Data obtained with
  the PBE0 functional.}
  \label{fgr:Firpicisomer}
  \end{center}
\end{figure}

We report our energetic analysis concerning ligand opening and
isomerization of FIrpic in Tab.~\ref{tddft_pbe0}. A schematic view of
the most relevant transition paths for Ir-ligand bond breaking is
depicted in Fig.~\ref{fgr:plotstatespbe0}, and a selected set of
conformations is showed in Fig.~\ref{fgr:Firpicopening} and Fig.~\ref{fgr:Firpicisomer}.

Let us begin with the closed conformation of the most stable isomer
(named \emph{closed}), which we take as a reference through the entire
discussion (see Fig.~\ref{fgr:firpic_closed}). Electronic excitation
has little effect on the relative geometrical arrangement of
the ligands, and the same holds true for internuclear distances of
bonds formed by iridium with F$_2$ppy moieties. As concerns the
picolinate fragment, excitation of the molecule to a triplet state
induces a small but noticeable elongation of the Ir-N bond (about 0.05~\AA),
while in the first excited singlet the Ir-O bond shortens by $\sim0.1$~\AA,
of by $\sim~0.01$~\AA\ in the first triplet state. Our results agree very
well with previous calculations reported in literature~\cite{Gu2008,Li2013}.

\begin{table}
\begin{center}
  \begin{tabular}{crrrr}
    \hline
    Geometry  & Ground state & Triplet & Exciton triplet & Exciton singlet\\
    \hline
    Closed & 0.000 & 2.722 & 2.682 & 2.692  \\
    N-pic TS\textsuperscript{\emph{a}} & 1.150 & 3.094 & 3.130 & 3.514  \\
    N-pic open & 0.444 & 2.846 & 2.882 & 3.245  \\
    O-pic TS & no state & no state & 3.800 & 3.986  \\
    O-pic open & no state & no state & 3.729 & 3.752  \\
    C-F$_2$ppy TS & 5.194 & 4.886 & 4.972 & 5.597  \\
    C-F$_2$ppy open & 5.144 & 4.792 & 4.794 & 4.873  \\
    N-F$_2$ppy TS (open $\rightarrow$ A) & no state & 3.102 & 3.176 & 3.687  \\
    N-F$_2$ppy open (A) & no state & 2.988 & 3.031 & 3.673  \\
    N-F$_2$ppy TS (A $\rightarrow$ B) & 2.446\textsuperscript{\emph{b}}  & 3.859 & 3.830 & 4.231  \\
    N-F$_2$ppy open (B) & 1.819 & 3.077 & 3.086 & 3.562  \\
    N-F$_2$ppy TS (B $\rightarrow$ isomer) & 2.586 & 3.308 & 3.299 & 3.602  \\
    \hline
  \end{tabular}\\
  
  \textsuperscript{\emph{a}} TS is an acronym for transition state.
  \textsuperscript{\emph{b}} The transition state in this case corresponds to the transition from the
  closed geometry to the state B.
  \caption{Relative energies (eV) of several FIrpic conformations
  with respect to the closed ground state minimum. All conformations are fully relaxed.
  Data obtained with the PBE0 functional.}
  \label{tddft_pbe0}
\end{center}
\end{table}

The bonds between Ir and the ligands have quite different strength,
and opening of the picolinate fragment is much more viable through
breaking of Ir-N as compared to Ir-O. When the molecule is in its
ground state (GS), this process requires 1.150~eV (111~kJ/mol), a value reduced to
0.822~eV (79~kJ/mol) when the first excited singlet is considered. When the
molecule is in the lowest triplet, a much smaller activation energy
is needed: 0.372~eV (36~kJ/mol) as estimated by single determinant (SD)
triplet computations, and 0.448~eV (43~kJ/mol) at TDDFT level. The metastable
open state is very stable in the GS, 0.706~eV below the transition
barrier. Furthermore, once N detaches from Ir, the two oxygen atoms
rearrange and both become ligands of the metal atom, whose coordination
remains essentially a distorted octahedron.

The situation is quite different as far as excited electronic states
are considered. First, open conformations obtained by Ir-N bond
breaking reside about 0.25~eV (24~kJ/mol) below the transition barrier. Second,
upon breaking of Ir-N FIrpic abandons the distorted octahedral
conformation of the closed molecule and becomes a trigonal bipyramid,
where Ir has just five ligands and the open pyridine fragment moves
far away from the bonding region with Ir. When the molecule is
electronically excited, the geometrical arrangement of fragments
in the transition state (TS) is similar to the corresponding open
minimum. The most significant difference comes from the torsion
angle between CO$_2$ and pyridine within the picolinate fragment,
which is very small either in closed and open conformations and
somewhat larger in TS. Its value ranges from 7\textdegree\, considering
the transition state of the electronic ground state to about
20\textdegree\, when electron excitations are considered.

Elongation of the Ir-O bond leads to no metastable conformation
in the ground state FIrpic molecule. As for the spin triplet, an
open state might not exist, as predicted by single-determinant computations,
or be very shallow, less than 0.1~eV below the barrier at the TDDFT level,
a value comparable to the thermal energy at 300~K. The opening barrier
is significantly higher in the excited singlet than in the triplet,
1.294~eV (124~kJ/mol) to be compared with 1.118~eV (107~kJ/mol), but the open state is markedly
deeper in the first case, 0.234~eV (22~kJ/mol) below the TS. In the open state
generated by elongation of Ir-O bonds, a relevant torsion between
the CO$_2$ and the pyridine fragments of picolinate occurs, and
reaches up to 50\textdegree. At the same time, pyridine rotates away
from the C-Ir-C plane by about 20\textdegree.

Let us now consider the opening of the F$_2$ppy ligand. The bond
with carbon is the strongest among the ones formed by the central
Ir atom. When the molecule is in its electronic ground state,
ring opening at Ir-C has an exceedingly high barrier, larger than
5~eV. Electron excitation into a singlet state reduces this value
to about 3~eV, and the activation energy further reduces to
slightly more than 2~eV in the first triplet excited state. The open
metastable state is quite shallow as far as the GS and
the triplet excitation are considered, while it is very deep (0.724~eV, 70~kJ/mol)
in the first excited singlet. As for the conformation, after
Ir-C breaking FIrpic evolves towards a square base pyramid. The open
F$_2$phenyl fragment presents a torsion angle with pyridine of
56\textdegree\, in the GS. This value is largely increased in the
excited states: 122\textdegree\, in the excited singlet,
124\textdegree\, in the exciton singlet and 138\textdegree\, in the
SD triplet. Transition states between open and closed Ir-C
conformations are characterized by a moderate distortion of the
square pyramid (the N-Ir-N bond angle involving F$_2$ppy fragments
slightly increases from about 110\textdegree\, to 120\textdegree),
while a strong reduction of the torsion between open F$_2$phenyl
group and pyridine occurs.

Opening of Ir-N bonds of F$_2$ppy requires to overcome energy
barriers comparable to the analogous bond of picolinate. Moreover,
when the molecule is electronically excited, this process represents
a feasible way towards isomerization of FIrpic, involving two
different intermediate states. Indeed, the elongation of Ir-N leads
to a first metastable conformation (labelled A in Table~\ref{tddft_pbe0})
which is not present in the GS molecule. In state A, both Ir-N bonds
of F$_2$phenyl increase their internuclear distance, about 0.4 and 0.2~\AA\,
for the opening and the closed ligands, respectively. The related N-Ir-N
bond angle markedly decreases from about 180\textdegree\, to about
145\textdegree\, (excited triplet) and 163\textdegree\, (excited singlet).
The excited triplet has a barrier smaller than 0.5~eV (48 kJ/mol), and a large
stability of the open states (about 0.5~eV below the barrier).
Conversely, the excited singlet has a large barrier, about 1~eV, and
a negligible stability range. Considering that no analogous metastable
state is found for the molecule in the singlet ground state, we argue
that intermediate state A is stabilized by a spin triplet electronic
configuration. Once the Ir-N bond elongates up to about 2.5~\AA\, to reach
conformation A, the system can evolve towards a second metastable
state, which has been found also when the molecule is in its electronic
ground state, and is labelled as B in Table~\ref{tddft_pbe0} and
Fig.~\ref{fgr:Firpicisomer}. Interestingly, in state B the relative
conformation of picolinate and closed F$_2$ppy fragment does not
correspond anymore to the starting closed FIrpic isomer, but closely
mimics the arrangement of a different isomer. The rearrangement of
the picolinate and the closed F$_2$ppy fragments occurring in state
B induces steric hindrance on the open F$_2$ppy fragment, which
counteracts either by rotating the open pyridine with respect to the
difluoro-phenyl group, as occurs in the ground state, or by rotating
the entire F$_2$ppy fragment towards a spatial region free of atoms,
as happens in the excited states.

Considering the GS molecule, the activation energy required to evolve
from the closed to the B metastable state is very high, 2.446~eV (236~kJ/mol).
However, when the system is electronically excited, energy barriers
from A to B are much smaller, less than 0.8~eV. Moreover, the B
conformation is very stable with respect to the A to B transition state,
about 0.5~eV below the top of the barrier. Transition states between
A and B are characterized by a relative arrangement between the
picolinate and the closed F$_2$ppy fragments half the way between
the initial and final closed isomers. Indeed, the relative N-Ir-N
angle measures about 140\textdegree, to be compared with 90\textdegree\,
in the minimum energy starting isomer, and 180\textdegree\, in
the final closed isomer. Once the open state labeled as B is formed,
the evolution towards the final closed isomer requires smaller
barriers than the previous steps: 0.767~eV (74~kJ/mol) in the GS, a value that
decreases to about 0.2~eV (19~kJ/mol) in the excited triplet and to 0.04~eV (4~kJ/mol) in
the excited singlet. Overall, isomerization along the described
path requires at least 2.5~eV (241~kJ/mol) when FIrpic is in the GS, but
when the molecule is in its first excited singlet the activation
energy required is 1~eV smaller, a value further reduced to about
1.1~eV (106~kJ/mol) when the molecule is excited to a triplet state.
Finally, we note that people have argued and shown, that different isomers
might display different non-radiative channels due to their complex excited
state PES.~\cite{Setzer2017,ArroligaRocha2018,Yao2018}

\subsection{Car-Parrinello simulations}
First, we performed ab-initio molecular dynamics (MD) runs of FIrpic in the
neutral, positively and negatively charged state. These runs revealed that
no rearrangements but thermal motion takes place, and that the ligands
around Ir retain the conformation of a distorted octahedra. This is
consistent with static computations, since in the neutral GS molecule
bond breaking involving Ir and its ligands requires activation energies
larger than 1~eV, a value much higher than vibrational energy at 1000~K.

\begin{figure}
  \begin{center}
  \includegraphics[width=8cm]{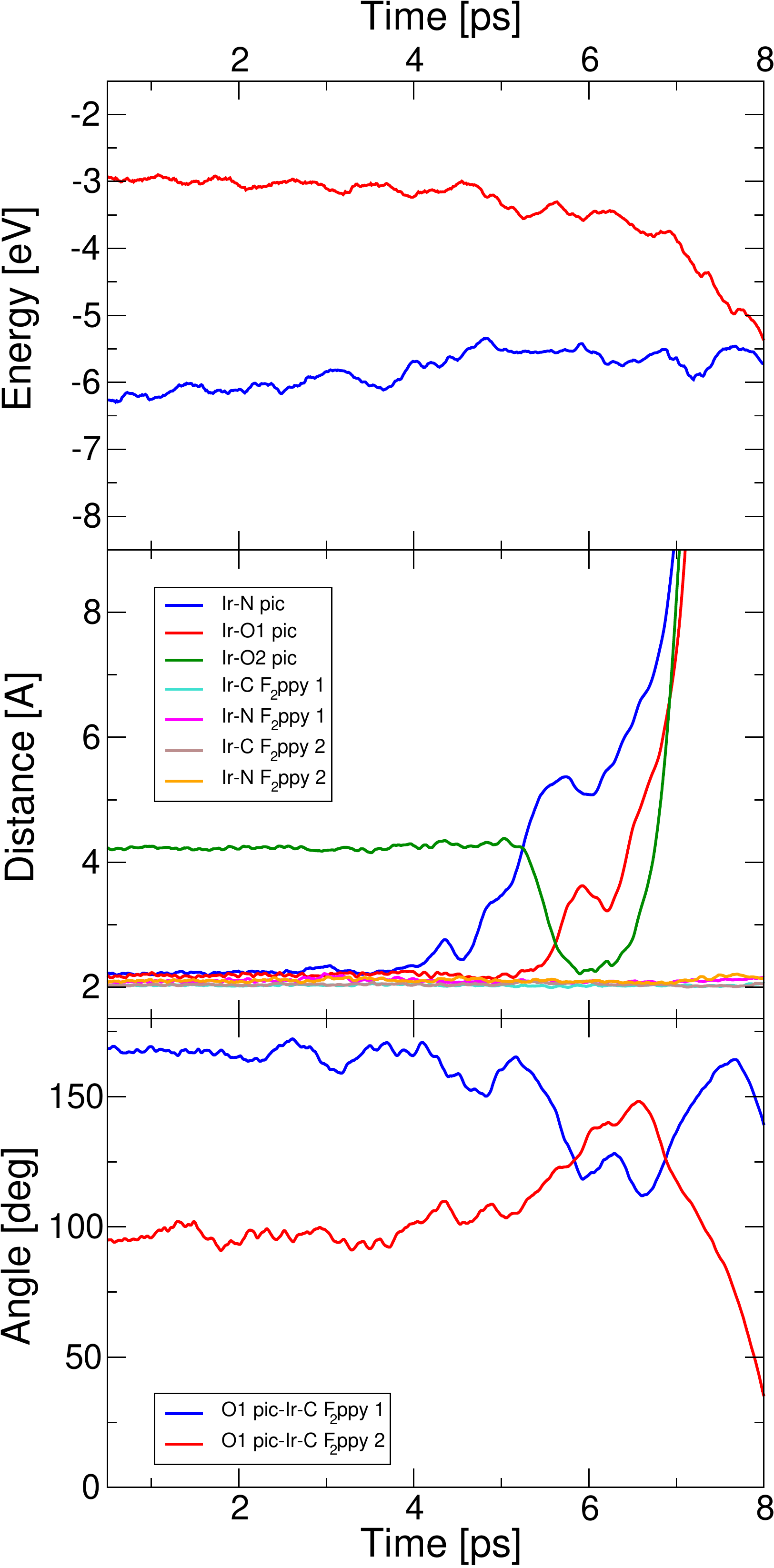}
  \caption{Dynamical evolution of energetic and structural parameters of
  FIrpic as extracted from CP Simulation I. Top panel: eigenvalues (eV) of
  the highest singly occupied molecular orbitals. Middle panel: internuclear
  distances (\AA) between iridium and the six ligands, plus the separation
  between Ir and the second oxygen of the picolinate fragment (O2).
  Bottom panel: angles (degrees) formed by the oxygen atom bonded to iridium
  (O1), the central metal atom, and the carbon atoms belonging to the F$_2$ppy
  fragment and bonded to Ir. We made the plots smoother and more readable
  by averaging values over 25 data points.}
  \label{fgr:Firpic_evolution}
  \end{center}
\end{figure}

\begin{figure}
  \begin{center}
  \includegraphics[width=8cm]{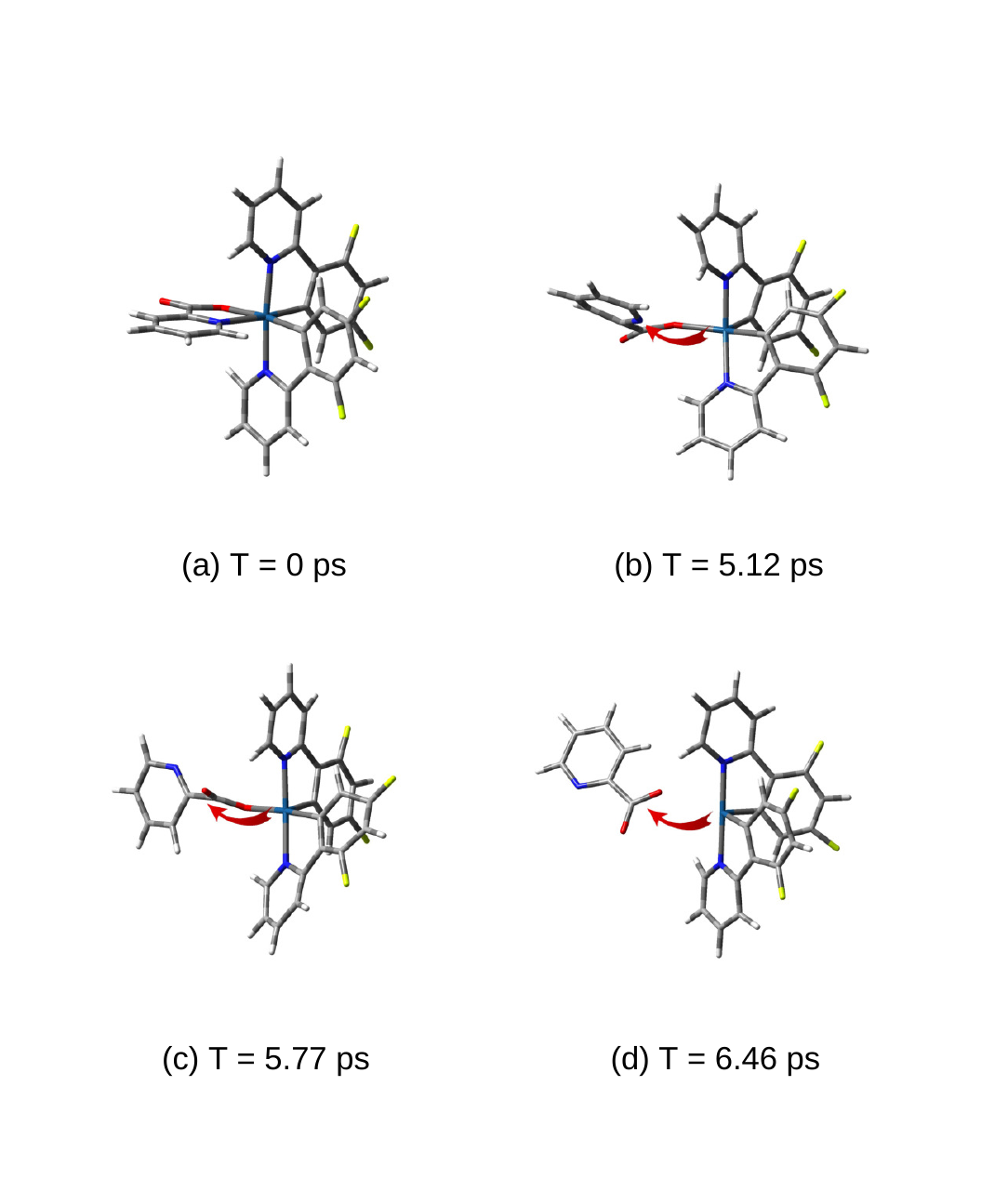}
  \caption{Selected snapshots of the CP dynamics of FIrpic in the first
  excited state. (a)~Starting closed conformation; (b)~The Ir-N bond
  elongates and nitrogen detaches from iridium; (c)~The oxygen atom
  bonded to iridium detaches and is substituted by the second
  carboxylate oxygen; (d)~The Ir-O bond elongates and the entire picolinate
  fragment abandons FIrpic, thus leading to irreversible degradation.}
  \label{fgr:FirpicCP}
  \end{center}
\end{figure}

The above picture radically changes when the non-spin-polarized exciton
is considered. Indeed, in two out of three MD runs of this systems,
we observed breaking of bonds between the metal atom and one ligand,
and even complete fragmentation of the molecule. In this section
we'll describe extensively just one of the three simulations
(from now on referred to as Simulation~I), while the other two
(Simulations~II and III) are only briefly reviewed. We report our results
on Simulation~I in Fig.~\ref{fgr:Firpic_evolution}, while the raw data
from the entire set of simulations are reported in the SI. In
Fig.~\ref{fgr:FirpicCP} we report four snapshots of the MD trajectory
extracted from the c-CPMD simulation of the exciton.

As can be seen in Fig.~\ref{fgr:Firpic_evolution}, the relevant structural
parameters (i.e. Ir-X bond lengths and X-Ir-X bond angles) remain quite
constant up to 4.05~ps, when the Ir-N bond involving the picolinate fragment
suddenly starts elongating. Once Ir-N is longer than $\sim$4~\AA, an
important change in the overall conformation of FIrpic takes place.
Indeed, as can be seen in the bottom panel of Fig.~\ref{fgr:Firpic_evolution},
the C-Ir-O bonding angles (where carbon atoms belong to the F$_2$ppy
fragments, and oxygen to picolinate) become similar to each another.
This is consistent with static computations where, upon Ir-N bond breaking
at the picolinate fragment, FIrpic abandons the octahedral conformation
and becomes a distorted trigonal bipyramid, characterized by almost equal
C-Ir-O angles.

At 5.20~ps the elongation of Ir-N and the rearrangement of the ligands
around the central metal atom induces a second and quite unexpected
effect on the conformation of FIrpic. Indeed, the oxygen atom not bonded
to Ir approaches the metal, and at 5.85~ps Ir-O shortens up to about
2~\AA, the typical bonding distance between iridium and oxygen. At the
same time lapse, the oxygen atom formerly bonded to Ir moves away from
the central metal atom reaching an internuclear distance of about 3.5~\AA,
where the bond with the metal is practically broken. According to
static computations, this process requires a very small activation
energy: 0.16~eV considering the single-determinant triplet, 0.11 and
0.14~eV at TDA level for the first singlet and triplet excited states,
respectively.

Then the molecule reaches a kind of steady state at about 5.95~ps, when
Ir-N measures about 5.5~\AA, and the nitrogen too distant from Ir in
order to form a chemical bond. At this time step the two oxygen atoms
of the picolinate fragment swap their positions as ligands of the central
metal atom, which forms just five bonds instead of the six present in the
starting closed conformation. Angles between carbon, iridium and oxygen
confirm that the molecule is closer to a trigonal bipyramid rather than
to an octahedron. Few picoseconds later, more precisely at 6.25~ps,
elongation of Ir-N takes place again and the same happens for both Ir-O
distances. The picolinate fragment abandons FIrpic.

We also note that the frontier (HOMO/LUMO) eigenvalues of FIrpic remain quite 
separated from one another during the MD run, until the picolinate fragment
is lost. This suggests that fast, non-radiative recombination is not likely to
occurr. Moreover, the driving mechanism of fragmentation originates from
breaking of Ir-N involving the picolinate fragment, which according to static
computations is the weakest bond among the six formed by the metal atom.

Simulation~II shows essentially the same mechanisms discussed above.
Ir-N opening at the picolinate fragment starts at 2.5~ps, and about 0.5~ps
later the two oxygen atoms swap as ligand of iridium. The short Ir-O
bond elongates up to about 4~\AA, and the oxygen not bonded to iridium
substitutes the former one as metal ligand. Quite interestingly, in
Simulation~II the two oxygens swap again their relative position
with respect to Ir before the fragmentation of FIrpic takes place.
This phenomenon manifests at 4.7, 5.2 and 6.0~ps, and no dissociation
of the picolinate fragment from iridium takes place until 7.8~ps.
Its driving force is also in this case a sudden elongation of Ir-N.
Differently from simulation~I, in this case the frontier eigenvalues
gets very close to each other even before fragmentation, which implies
that fast non-radiative recombination should not be ruled out.

In simulation III we observed a quite different behavior with respect
to the other two. First of all, elongation and breaking of Ir-N is
not followed by the exchange between the two oxygens as Ir ligands.
Indeed, they swap position shortly twice, at 9.0 and 9.7~ps, quite
a long time after breaking of Ir-N, which occurs at 3.1~ps. Second,
no fragmentation is observed, and FIrpic is left, for quite a long time
(7.7~ps), with just five ligands around Ir instead of six.
Here, fragmentation does not appears to be a necessary consequence
of ring opening. Third, as observed in simulation II the frontier
eigenvalues approach each other after breaking of Ir-N.

A final comment concerning all MD three runs on FIrpic is mandatory.
Once Ir-N is broken, the unbound pyridine group of picolinate should
be free to rotate, as no relevant steric hindrance should be
encountered. As a matter of fact, the torsion of this group with
respect to the carboxyl group is somewhat correlated to the relative
position of the oxygen atoms with respect to Ir, such as the nitrogen
atom gets as close as possible to the oxygen bound to iridium.

After the dissociation of the picolinate we also observed detachment
of a CO$_2$ molecule from the picolinate (reported in literature),
leaving a C$_5$H$_4$N$^+$ fragment. However the closing of the
HOMO-LUMO gap, the Car-Parrinello method is not suitable to describe
non-adiabatic effects, hence our observation of the CO$_2$ molecule
evolution in a very short timescale (few ps) could be an artifact of
the method.

Before closing, we would like to highlight two points: the first is
that the rather high temperature in simulations (1000~K $\simeq$
0.083~eV) was used only to accelerate the dynamics. We verified
that in the neutral state at 1000~K, FIrpic remains intact for the
duration of the simulation. Thus, the simulation temperature has a
little influence on the electronic excitations, and thermal
dissociation might occur on longer time-scales.

The second is that our CPMD simulations are not meant to reproduce a
statistical ensemble in thermal equilibrium. Rather, they represent
a tool to generate and explore possible degradation paths. Indeed we
cannot extract lifetimes from our simulations, as non-adiabatic effects
are not described accurately by the method. The lack of statistical
sampling means that the method cannot predict the relative probability
of each single degradation event.

\section{Conclusions}
In this paper we unraveled the degradation mechanism of FIrpic by
means of theoretical computations, and we also identified a possible
pathway for its isomerization. Ab-initio molecular dynamics simulations
predict the no degradation occurs when the molecule is in the
ground state, and the same holds true when FIrpic is positively
or negatively charged. Static computations confirm this outcome,
since bond breaking between the central metal atom and its ligands
requires activation energies ranging from 1 to 5~eV depending on
the bond considered. 

The picture radically changes when an excited electronic configuration
is considered. All bonds formed by Ir become weaker, and the Ir-N
ones become the loosest. Static computations indicate that when
the molecule is excited in its triplet state, breaking of the
Ir-N at the picolinate and the F$_2$ppy fragments requires an
activation energy of about 0.5~eV, while about 1~eV is needed when
in the first excited singlet. Interestingly, in our Car-Parrinello
simulations of the excited molecule we invariably observed
Ir-N breaking at the picolinate fragment, followed by formation
of a metastable open intermediate with a distorted trigonal
bipyramidal arrangement of the ligands around iridium. Molecular
dynamics simulations also showed that after Ir-N opening, the
two oxygens of picolinate reversibly swap as ligands of iridium,
which thereafter remains five-fold coordinated.

This process pave the way towards fragmentation of FIrpic, which
loses the picolinate fragment in two simulations out of three,
a result in nice agreement with the general consensus on this subject.
As a further confirmation of this picture, in a parallel experimental
investigation (Paper~II~\cite{paperII}) we proved that the loss of picolinate is indeed
the key step in the degradation of FIrpic based devices. Furhtermore,
we discovered that the resulting Ir(F$_2$ppy)\textsuperscript{+} species
favorably bind to BPhen, a material widely used as electron transport
layer in OLED devices. Remarkably, the electron emission spectrum of
[Ir(F$_2$ppy)$_2$BPhen]\textsuperscript{+} is significantly redshifted
with respect to the unaged FIrpic. This outcome has been confirmed
by static computations conducted on this molecule at the same
level of theory adopted for FIrpic.

Another fallout of the aging of FIrpic based devices is isomerization.
Contrary to previous computational investigations, we proved that
isomerization cannot be afforded simply by rotating the moieties
bonded to iridium. Indeed, the first step is necessarily the opening
of one Ir-ligand bond, followed by a quite complex conformational
rearrangement of the entire molecule. The overall energy barrier
for the isomerization of the ground state FIrpic molecule is
about 2.5~eV, a value reduced to about 1~eV when the first excited
singlet and triplet states are considered.

In conclusion, by means of state of the art computations we provided
a rationale for the degradation and isomerization mechanisms of
FIrpic, which go through the same initial steps. Consistently with
experimental evidences, it comes out that at real operating temperatures
both processes are viable only when the molecule is electronically
excited, and more so when the first excited triplet is considered.
Overall, our results provide a solid physical basis to establish
the relation between the excitation/de-excitation processes and
the degradation of FIrpic in OLED devices.

\begin{acknowledgement}
We acknowledge the CINECA award under the ISCRA initiative, for the
availability of high performance computing resources and support
(Grants Nr. HP10CH6GBY, HP10CKPLJY and HP10CJYU54). This project
was funded by the GRO 2014 and 2015 program of Samsung Advanced Institute
of Technology (SAIT). We thank Woon-jon Son, Dongseon Lee and
Rocco Martinazzo for useful discussions.
\end{acknowledgement}

\begin{suppinfo}
Additional results from static computations performed at the B3LYP
level and c-CPMD simulations results for three different trajectories
simulating the exciton. 
\end{suppinfo}

\bibliography{OLED}

\end{document}